\documentstyle[preprint,aps,epsfig]{revtex}

\begin{document}
\draft

\title{The sign of temperature inhomogeneities deduced from time-distance 
helioseismology}

\author{M. Br\"uggen} 
\address{Max-Planck-Institut f\"ur Astrophysik,
85740 Garching, Munich, Germany; and Churchill College, Cambridge, UK}
\author{H.C. Spruit} 
\address{Max-Planck-Institut f\"ur Astrophysik,
85740 Garching, Munich, Germany}

\date{\today}
\maketitle

\begin{abstract}

Inhomogeneities in wave propagation conditions near and below the
solar surface have been detected by means of time-distance
helioseismology.  Here we calculate the effect of temperature
inhomogeneities on the travel times of sound waves. A temperature
increase, e.g. in active regions, not only increases the sound speed
but also lengthens the path along which the wave travels because the
expansion of the heated layers shifts the upper turning of the waves
upward. Using a ray tracing approximation we find that in many cases
the net effect of a temperature enhancement is an {\it increase} of
the travel times.  We argue that the reduced travel times that are
observed are caused by a combination of magnetic fields in the active
region and {\it reduced} subsurface temperatures. Such a reduction may
be related to the increased radiative energy loss from small magnetic
flux tubes.

\end{abstract}

\section{Introduction}

In recent years it has become possible to measure the travel times of
acoustic waves travelling through the outer layers of the Sun through
`time-distance helioseismology'. These travel times are used to infer
information about the sub-surface structure of the Sun and have
revealed inhomogeneities in the wave propagation conditions. (Duvall
et al. 1993, Duvall 1995, D'Silva \& Duvall 1995, Duvall 1997, Duvall
et al. 1998). These inhomogeneities indicate shorter sound-wave travel
times and have been associated with active regions.\\

Slight shifts in global p-mode frequencies in the course of a solar
cycle have also been detected (Woodard \& Libbrecht 1993, Dziembowski
et al. 2000). As in the case of the time-distance measurements,
they have been found to be closely related to magnetic activity
(Dziembowski et al. 2000), and indicate shorter travel times in
the outermost layers of the convection zone. A common interpretation
for both effects is that the sound speed in active regions is somewhat
higher because of an increase in temperature. However, it was shown by
Goldreich et al. (1991) that, though intuitively appealing, such
a temperature increase is unlikely to be the cause of the mode
frequency shifts. Contrary to expectation, a higher temperature causes
{\em longer} travel times, because a temperature increase causes a
slight expansion of the envelope. The path length increase caused by
this expansion dominates over the increase in sound speed since the
expansion is proportional to the temperature $T$, while the sound
speed increases only as $T^{1/2}$. Instead of a temperature increase,
Goldreich et al. propose that most of the effect is due to the
photospheric magnetic field. A magnetic field increases the stiffness
of the gas as experienced by pressure waves.\\

The expansion argument would also affect the interpretation of
travel-time anomalies discovered by time-distance helioseismology. The
purpose of this paper is to verify to what extent the expansion of the
envelope associated with a temperature rise increases the wave travel
times. In Sec. 2 we show that in most cases a temperature enhancement
does in fact lead to {\em longer} travel times. To explain shorter
travel times one might then invoke magnetic fields.
Alternatively, the subsurface temperatures could be slightly {\em
lower} in the active regions, the opposite of what is concluded in
most analyses of time-distance measurements.  These possibilities are
discussed briefly in Sec. 3.

\section{Model}

One would like to investigate the effects of temperature
inhomogeneities without making too many assumptions about their
origin. On the other hand, for actual wave propagation calculations a
well defined equilibrium model is needed. A simple model is a {\em
geostrophic} one, in which the pressure balance within the
inhomogeneity is attained through Coriolis forces associated with the
solar rotation. This balance is probably valid on large scales (of the
order of an active region). On smaller scales, this is probably not a
good model; see Sec.\ref{circs} for a detailed discussion.

For definiteness in discussing the model, we regard a region of
increased temperatures. Since the effects are nearly linear in $\delta
T$, the results apply equally with opposite sign to regions of lower
than average sub-surface temperatures.

\subsection{Equilibrium model}

In a plane-parallel envelope in hydrostatic equilibrium under constant
gravitational acceleration $g$, the pressure, $p$, satisfies

\begin{equation}
\frac{dp}{dz}=g\rho ,
\end{equation}
where $z$ is the depth beneath some reference layer. Assuming a
polytropic relation between $p$ and density, $\rho$, i.e.

\begin{equation}
p = K\rho^{1+1/n} ,
\end{equation}
one finds that the pressure varies with depth as

\begin{equation}
p = p_0\left (\frac{z}{z_0} \right )^{n+1}
\end{equation}
and similarly

\begin{equation}
\rho = \rho_0\left (\frac{z}{z_0} \right )^n ,
\end{equation}
where $z_0$ denotes the depth of some reference layer from where the
polytropic layer extends downward. Above $z_0$ one may, for example,
want to match the polytrope onto an isothermal atmosphere. Here we 
only consider rays that lie completely within the polytrope.

The sound speed is thus
\begin{equation}
c^2 = \frac{\gamma g}{n+1} z ,
\end{equation}
where $\gamma$ is the ratio of the specific heats.
The acoustic cut-off frequency is approximately given by

\begin{equation}
\omega_{\rm c} = \frac{c}{2H} = \left (\frac{\gamma g}{n+1} \right 
)^{1/2}\frac{n}{2}z^{-1/2},
\end{equation}
where $H$ is the density scale height defined as

\begin{equation}
H = \left (\frac{{\rm d}\ln\rho}{{\rm d}z} \right )^{-1} = z/n .
\end{equation}
This yields a depth of the upper turning point of
\begin{equation}
z_{\rm t} = \left (\frac{n}{2\omega}\right )^2 \frac{\gamma g}{n+1} 
.\label{zt}
\end{equation}
Assuming the gas to be ideal and the ionization to be constant, the
temperature is given by

\begin{equation}
T = \frac{\mu p}{R\rho} = \frac{\mu g}{R(n+1)} z ,
\end{equation}
where $\mu$ is the mean molecular mass and $R$ the gas constant.
Thus, the temperature profile is determined by the polytropic
index $n$. 

The entropy is given by

\begin{equation}
S\sim\ln p/\rho^\gamma= (1+1/n-\gamma)\ln\rho + {\rm cst.},
\end{equation}
where an ideal gas equation of state has been assumed. Since the stratification 
of the convection zone is close to adiabatic, the values for the polytropic index 
and $\gamma$ are related by
\begin{equation} \gamma\approx 1+1/n .\end{equation}

\subsection{Time-distance calculations}

We now investigate the travel times of those waves that enter a
column of hotter material, are then reflected near the surface and
subsequently leave the hotter region again.  We treat this problem in
two dimensions (i.e. a slab geometry); we ignore the effect of the
advection of the waves, assuming that to equal parts
the waves travel in and out of the inhomogeneity and that therefore
the net effect of advection vanishes to first order. The flows may
cause second-order effects ($\sim v^2$, and independent of the
direction of the flow) on the travel times. This is discussed in
section \ref{circs}.\\

In the spirit of the JWKB approximation we will regard the sound waves
as locally plane (see Gough 1993). In this approximation, which is
commonly made in local helioseismology, the waves are assumed to
follow rays that obey the laws of geometrical acoustics.  For
simplicity, we will regard the medium as plane-parallel. \\

We let the ray traverse a column in which the
temperature profile is raised to $T_1(z)$. This will be achieved using
two very simple models. In model 1 the temperature in the hotter
column is raised by lowering the polytropic index to $n_1<n_0$
(subscript 0 shall denote the corresponding values of the `normal'
Sun). The temperature difference in this model is thus proportional
to the depth $z$, and its depth is assumed infinite.\\

Model 2 mimics some proposed models for temperature enhancements
associated with active regions (Kuhn and Stein 1996). In these models
the source is assumed to be an entropy increase located at some depth
$D$, the effect of which extends to the surface. This can be
incorporated in a model with a higher, but still depth-independent
entropy $S$. In a polytropic model, this corresponds to an increased
value for the polytropic constant $K$. If this model is in lateral
pressure balance at the source depth $D$, vertical equilibrium implies
a vertical shift, such that the top of the polytrope is at some depth
$z_1<0$. If $K_1$ is the polytropic constant of this model, one finds
that
\begin{equation}
\left ({K_1\over K_0}\right )^{-{n\over n+1}}=1+{z_1\over D}.\label{pol}
\end{equation}
For small entropy changes $\delta K=K_1-K_0$ we have
\begin{equation} {z_1\over D}=-{n\over n+1}{\delta K\over K_0}.\end{equation}
The vertical shift is thus proportional to the depth of the source. In
the example shown in Fig. \ref{fig.3} we have chosen a depth of $D=21$
Mm and $\delta K\over K_0$ was chosen to be $6\cdot 10^{-3}$, so
that $z_1\approx 120$ km.

In the hotter column the upper turning point is raised over the one in
the adjacent colder medium. In case 1, $n_1 < n_0$, which decreases
the depth $z_{\rm t}$ of the upper turning point (see Eq.(\ref{zt}))
and in case 2 the upper turning point is shifted upward due to
expansion.\\

The ray equations follow from the dispersion relation, which can be
written as

\begin{equation}
k^2=k_{\rm v}^2+k_{\rm h}^2=\frac{\omega^2-\omega_{\rm c}^2}{c^2} ,
\end{equation}
where $k_{\rm v}$ and $k_{\rm h}$ are the vertical and horizontal
components of the wavevector ${\bf k}$.\\

The advantage of using a polytropic approximation (aside from the fact
that the approximation is quite good for the upper layers of the Sun)
is that the ray equations can be written down analytically. If
$x$ is the horizontal coordinate in the plane of the wave and assuming
that $k_x^2+k_z^2=\omega^2/c^2$ (i.e. ignoring $\omega_{\rm c}$), one
can write

\begin{eqnarray}
x & = 
&\int\frac{dx}{dz}dz=\int\frac{k_x}{k_z}dz=\int\frac{k_x}{\sqrt{\omega^
2/c^2-k_x^2}}dz = \int\frac{dz}{\sqrt{a/z-1}}\nonumber \\
& = & a\left [\sin^{-1}(z/a)^{1/2}-(z/a)^{1/2}(1-z/a)^{1/2}\right 
],\label{4.50}
\end{eqnarray}
where $a$ is the depth of the lower turning point given by
$a=\omega^2/c_0^2k_x^2$.
The time taken for this traverse is given by the integral along the ray

\begin{eqnarray}
\tau & = &\int\frac{k}{\omega}ds=c_0^{-1}\int\ z^{-1/2}(1-z/a)^{-1/2}\ 
dz \nonumber\\
& = & c_0^{-1}a^{1/2}\sin^{-1}(z/a)^{1/2} \ .
\end{eqnarray}
When the cut-off frequency is included in the ray equations, the
expressions become a bit more complicated, but the integrals can still
be solved analytically.\\

\begin{figure}[htp]
\centerline{
\psfig{figure=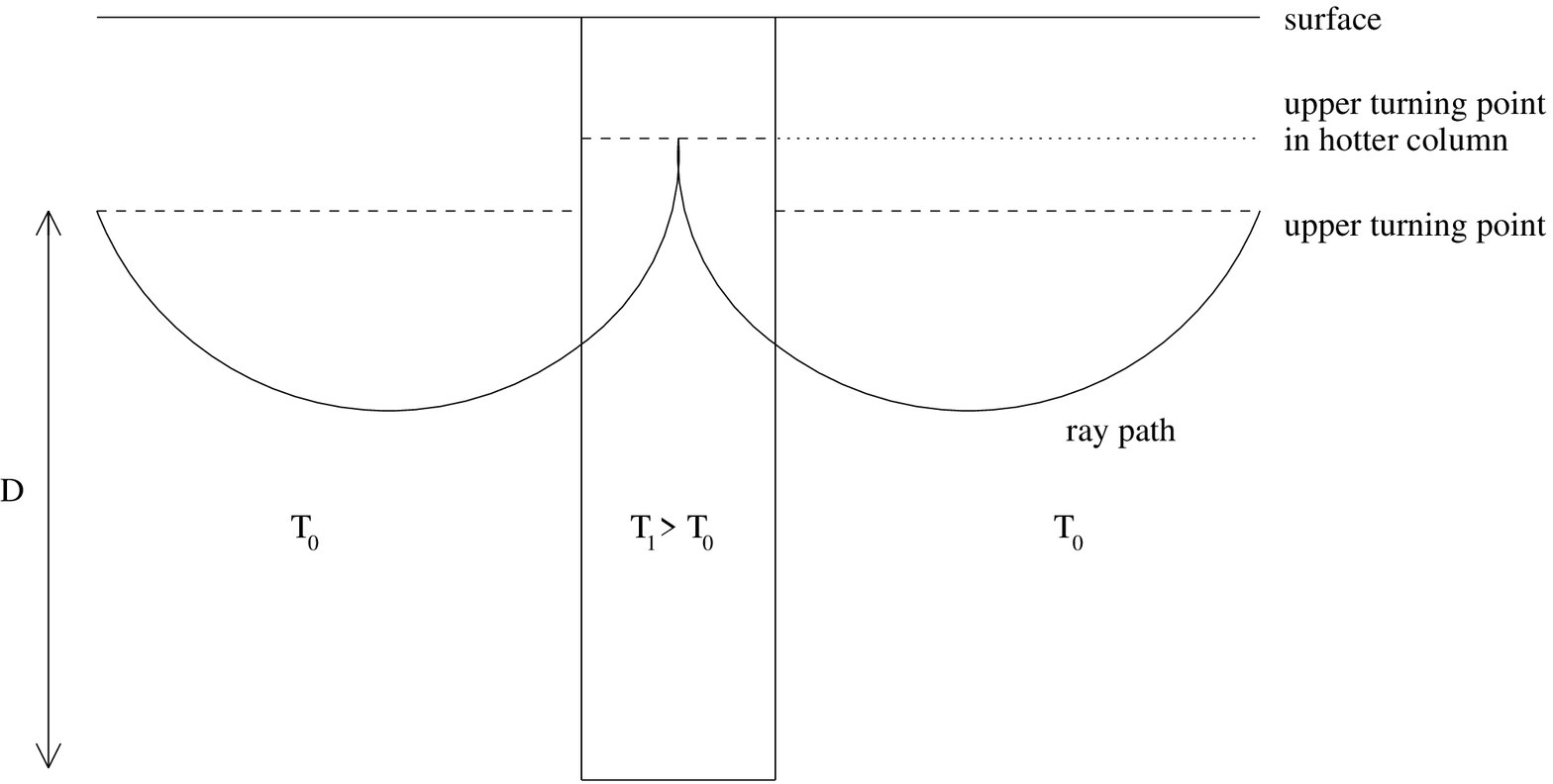,width=13.0cm,angle=0}}
\caption{Sketch of a ray which traverses and is reflected in a hot region.}\label{fig.2}
\end{figure}

\begin{figure}[htp]
\centerline{
\epsfig{figure=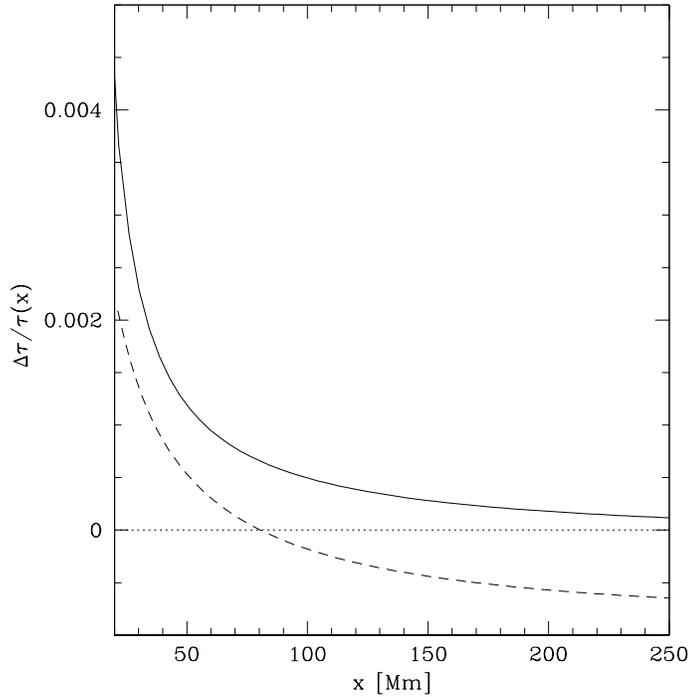,width=10.0cm}}
\caption{Relative travel-time differences between the homogeneous and
an inhomogeneous Sun as a function of interskip distance. The dashed
line shows the result from our model 1 with polytropic indices of
$n_0=3$ and $n=2.7$ inside the inhomogeneity which has got a presumed
width of $w=3.5$ Mm. The solid line shows the results from our model
2, in which the entropy has been increased down to a width of $D=20$
Mm.  ($\delta K/K_0=6\cdot 10^{-3}$, $w=7$ Mm)}\label{fig.3}
\end{figure}

\begin{figure}[htp]
\centerline{
\epsfig{figure=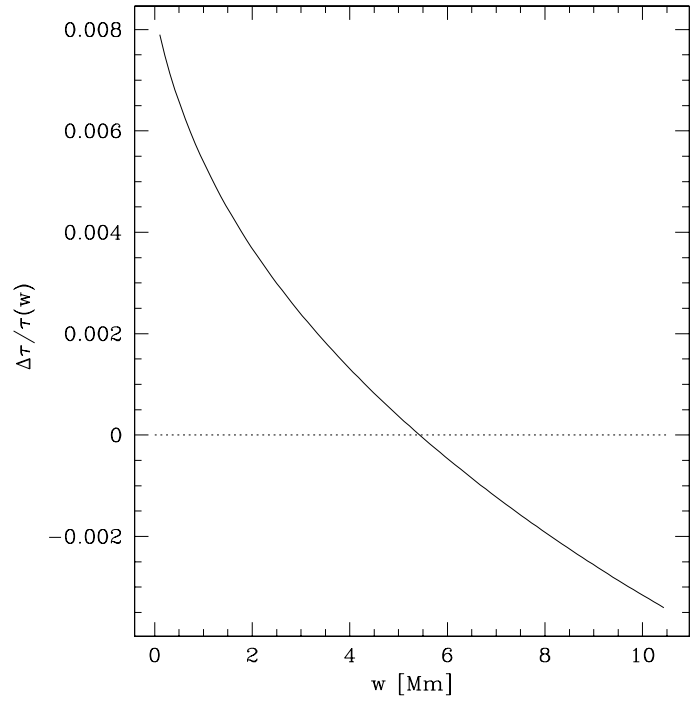,width=10.0cm}}
\caption{Relative travel-time difference between the homogeneous and an
inhomogeneous Sun for our model 1 as a function of the width of the
hot column for a ray with an interskip distance of 7 Mm.}\label{fig.4}
\end{figure}

At the vertical interface between the two regions of different
temperatures, the rays are refracted. We are using the approximations
of geometrical acoustics and calculate the angle of refraction by
Snell's law. Then we compare the travel times in the homogeneous Sun
with the corresponding times in the scenario depicted in
Fig. \ref{fig.2}. In all cases, we consider rays which bounce
(i.e. have their upper turning point) inside the inhomogeneity (see
Fig. \ref{fig.2}). In Fig. \ref{fig.3} we have plotted the relative
travel time difference $\delta\tau /\tau = (\tau_{\rm inhom}-\tau_{\rm
hom})/\tau_{\rm hom}$ (between rays in a homogeneous and inhomogeneous
Sun) for rays of different interskip distances (and depths) for a
fixed width of the hotter region in models 1 and 2. In
Fig. \ref{fig.4} $\delta\tau /\tau$ is shown for a ray of fixed depth
but as a function of the width of the hotter region.\\

Fig. \ref{fig.3} shows that the travel-time can be positive, i.e. that
the waves can take longer in the presence of the hot column. This
implies that the hot column retards the waves, because the effect of
the raised upper turning point outweighs the increased sound speed in
the hotter column. However, $\delta\tau /\tau$ decreases with
increasing depth of the ray, since the waves spend more time in the
hotter (and therefore faster) region whereas the effect of the raised
upper turning point depends very little on the depth of the rays
(since the rays are almost vertical near the upper turning point).\\

In Fig. \ref{fig.4} one can note that the travel-time is positive as
long as the width of the hot region is small compared to the length of
the ray.  But $\delta\tau /\tau$ decreases with increasing $w$ because
the sound speed is higher in the hotter region, and this
eventually dominates the effect of the lengthening of the ray through
the raised upper turning point: the travel-time difference becomes
negative.

\subsection{Horizontal equilibrium of a temperature inhomogeneity}
\label{circs}

\begin{figure}
\centerline{
\psfig{figure=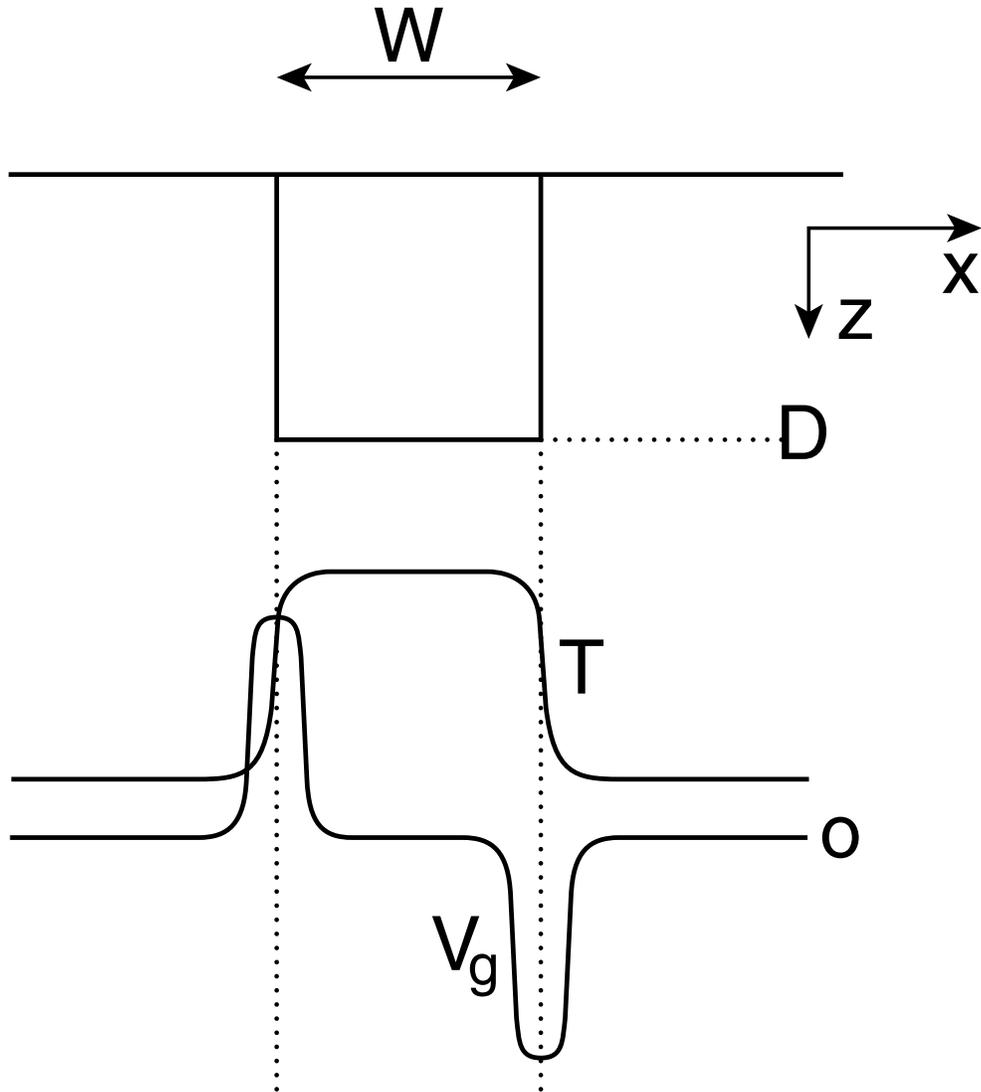,width=13.0cm,angle=0}}
\caption{Schematic picture showing a model of geostrophic
equilibrium of a column of enhanced temperature. The solid line
indicates the azimuthal flow speed and the dashed line the temperature
distribution.}\label{fig.1}
\end{figure}

So far we have evaded the question about what restores the horizontal
pressure equilibrium.  The difference in temperature between the
inhomogeneity and its surroundings causes a difference in gas pressure
that is subject to adjustments on the short hydrodynamic time scale.
Consider a temperature enhancement in a patch (an active region, say)
at colatitude $\theta$, extending below the surface as a column of
width $L$ (Fig. \ref{fig.1}). In geostrophic balance, the pressure
excess $\delta p$ is balanced by the Coriolis force acting on a flow
$\bf v$ around the column, i.e.
\begin{equation}
(\nabla\delta p)_{\rm h}=2(\rho {\bf v}\times \Omega)_{\rm h}, \label{cor}
\end{equation}
where the subscript ${\rm h}$ indicates the horizontal components.

The flow is concentrated at the boundary of the column, where the
pressure gradients are greatest. Inside the column, the flow
vanishes and the excess pressure $\delta p$ is just given by
hydrostatic equilibrium,
\begin{equation} 
p_r\delta p=-g\delta\rho.
\end{equation} 
The column is assumed to extend down to some depth $D$ where $\delta
p=0$. Below this depth, the temperature excess vanishes. The flow
speed is then maximal at the surface, and vanishes at depth $D$.\\

In the preceding paragraph, we have proposed a geostrophic balance for
the pressure changes. For relatively small inhomogeneities the effects
of rotation are small, and a geostrophic balance is not
realistic. Instead one could consider upwellings and downdrafts. If
there is a source of heat at some depth below the surface, a
circulation is set up, with upwelling above the heat source. This flow
is driven by the higher gas pressure inside the rising column, and its
velocity $\bf v$ is such that the dynamic pressure of the flow, $\rho
{\bf v}\cdot\nabla{\bf v}\approx \rho v^2/r_c$, balances the pressure
difference ($r_c$ is the gradient length scale of the velocity).\\

The flow advects the p-modes (to first order in $v$) but also has a
second-order effect on their propagation speed. The second-order
effect is proportional to, and of the same order of magnitude as the
temperature increase and likely to give a positive contribution to the
sound speed. This is because flows with vorticity speed up on
compression.  In the previous section we neglected the contribution to
the sound speed which is provided by the compressibility of the
flows. Its effect is to increase the sound speed, but the extent of
the effect is difficult to model.\\

\begin{figure}[htp]
\centerline{
\epsfig{figure=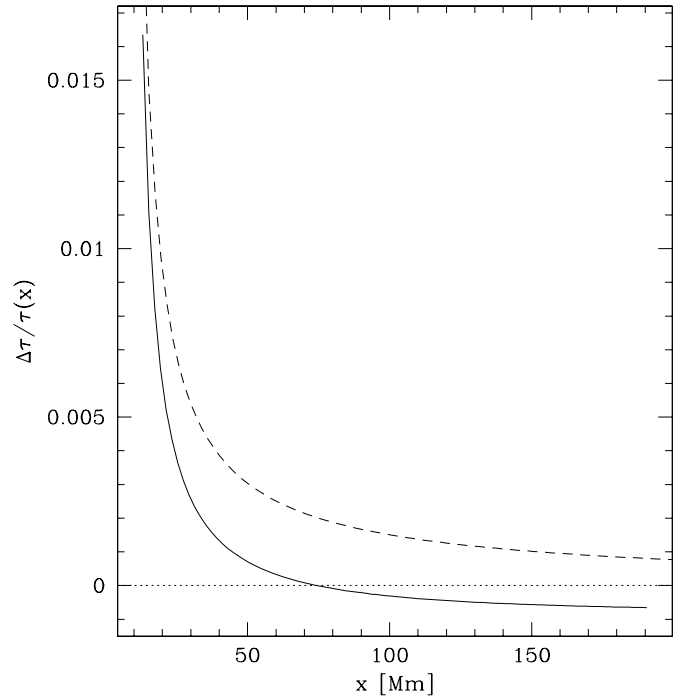,width=10.0cm}}
\caption{Relative travel-time difference for model 2 ($\delta
K/K_0=6\cdot 10^{-3}$, $w=7$ Mm, $D=40$ Mm) but with $\alpha = 1.01$
(see text). The dashed line shows the result with $\alpha = 1.00$.}\label{fig.5}
\end{figure}

For a qualitative estimate, we assume the effect of the flow on the
waves to be {\em local} and {\em isotropic}. In this case, the effect
on wave propagation would be a mere increase of the propagation speed.
Thus, we repeat the calculations presented above with an increased
sound velocity inside the hotter region:
\begin{equation} 
\delta c/c=\alpha {1\over 2}\delta T/T,
\end{equation}
where $\alpha$ is a factor of order unity. In the absence of flows, we
would have $\alpha=1$.  Fig. \ref{fig.5} shows the same case as
depicted in Fig. \ref{fig.3} only with $\alpha = 1.01$.  As expected,
the relative time delay is smaller than in the absence of flows and
even for a slight relative increase of the sound speed by 1 \% the
time delay quickly becomes negative. Therefore, one will have to know
the properties of the turbulent medium very accurately before one can
definitly predict its effect on the travel times.\\

In the above, the flows were presumed to have no effect on the
position of the upper turning point. In reality, a `turbulent'
pressure increase due to small scale flows expands the region
vertically. This would raise the upper turning point levels, and
increase the travel times. In the case of granulation flows, this
effect is the main contribution to the p-mode frequency anomalies
associated with the outer envelope (Rosenthal, et al. 1999). By leaving
it out, we are probably underestimating the travel time increase
(decrease) in hotter (cooler) regions.

\section{Summary and discussion}

Time-distance helioseismology indicates the presence of subsurface
inhomogeneities. In order to map these inhomogeneities quantitatively,
a propagation model is needed. In most analyses it is assumed that
shorter travel times correspond to higher propagation speeds. While
this is correct if the inhomogeneities are due to a vertical
magnetic field, the most obvious possibility, i.e. a change in
temperature, requires careful treatment. This is because a temperature
change has two side effects in addition to a change in propagation
speed. Since hot gas is less dense, the resulting positive buoyancy
causes the envelope to expand vertically in regions of higher
temperature. Secondly, the resulting horizontal imbalance also sets up
a circulation flow. Both have effects on acoustic wave propagation.\\

We have analysed here the effects of a temperature increase in a model
in which horizontal imbalance is compensated geostrophically by a
circulation (i.e. by Coriolis forces acting on a horizontal flow, as
in high- and low-pressure systems in the Earth's atmosphere). We find
that if temperatures are increased, but vertical expansion is ignored,
the changes in travel time as seen in time-distance measurements can
be of either sign, depending on the wavenumber and the inter-skip
distance. This is because the increased propagation speed is offset in
part by the fact that the upper turning point of the waves is higher
in a hotter model.\\

This effect is enhanced if vertical expansion of the perturbed model
due to vertical pressure balance is taken into account. We have
calculated this in a model in which the entropy has been increased
locally (corresponding to an increase in the polytropic constant
$K$). The results show that the travel times are {\it increased} by a
temperature enhancement, as long as the horizontal extent of the
inhomogeneity is not too large. We found it difficult to obtain time
delays for inhomogeneities that have widths of more than about 10 Mm
without making unrealistic assumptions.  If one takes into account the
effect of turbulence onto the sound speed inside the inhomogeneity,
one finds that the sign of the time delay can change. This implies
that predictions become sensitive to parametrizations of poorly known
turbulent flows and therefore less robust (see Sec. \ref{circs}).\\

The effects of temperature enhancements have been considered before 
by Kuhn and Stein (1996), whose conclusions differ from ours. 
By a numerical convection simulation, these authors calculate the effect of
a temperature increase applied at a depth of 2.6 Mm on a part of the
lower boundary. This increase causes both a circulation and a vertical
expansion of the model above the hotter boundary region. The
temperature change as a function of depth in their figure 2 mimics the
{\it vertical gradient} of the unperturbed model. The vertical shift
implied by the figure is about 10 km at the photosphere. This can be
compared with a value of roughly 12 km expected from approximate
vertical hydrostatic balance for an entropy change as applied by the
authors. The authors then perform a ray tracing calculation similar to
ours, and find {\it reduced} travel times.\\

The cause of the disagreement can be traced to the treatment by Kuhn
and Stein of the layers below 2.6 Mm, i.e. at depths not covered by
the numerical simulation. In their ray tracing calculations, they
assume higher temperatures, where $\delta T/T$ declines linearly from
0.006 at 2.6 Mm to 0.003 at a depth of 50 Mm. The vertical expansion
of the model, however, is based only on the 2.6 Mm layer included in
the simulation. The expansion is approximately proportional to the
depth over which the increased temperatures extend (see
Eqn. \ref{pol}). On the one hand, the convection simulation by Kuhn
and Stein demonstrates the vertical expansion effect, but on the other
hand their ray tracing calculation underestimates its effect on
travel times by a large factor.\\

Kuhn and Stein do not specify the cause of their assumed temperature
enhancement at a depth of 50 Mm, but suggest a source in those layers
where the magnetic field of the solar cycle is produced, near the base
of the convection zone ($z\sim 200$ Mm) (see also Kuhn, Libbrecht and
Dicke 1988). If this were the case, for a given temperature increase
at the surface, the vertical expansion effect would be four times
larger than for an assumed depth of 50 Mm. The travel time increase by
the vertical expansion would then certainly dominate over the
reduction due to the higher sound speed for all time-distance
measurements published so far (which reach depths of the order of 20
Mm).

\subsection{Lower temperatures in active regions?}

The shorter sound travel times in active regions found by
time-distance seismology are consistent with the increase of mode
frequencies with magnetic activity found by Woodard et al. (1993) and
Dziembowski et al. (2000). As noted by Goldreich et al. (1991), the
increased mode frequencies are not consistent with increased
temperatures, for the same reason as in our time-distance
calculations. Instead, Goldreich et al. suggest that the magnetic
field of active regions causes the increase in propagation speeds. For
the vertical magnetic fields seen near the surface, there would be no
associated vertical expansion of the envelope. While this explanation
is consistent with the data available then, it is no longer compatible
with recent data. This is because a magnetic change in propagation
speed is confined to a thin layer near the surface (unless very large
magnetic fields are assumed at a depth of 10--20 Mm). The mode
frequency changes measured by Woodard et al. and  very accurately
by Dziembowski et al. (2000) (with MDI) show that while most of the
effect is concentrated near the surface, there are also significant
changes at depths of 10 Mm. This would require field strengths of the
order 20 kG covering large fractions of the surface.\\

Instead of getting involved in a discussion about the difficulties that
such large unobserved magnetic fields would cause, we suggest here a
more radical and simple solution. One could infer that the subsurface
temperatures in active regions, in spite of the increased emission at
the surface, are in fact {\it reduced}.\\

If the effect of the small scale magnetic fields were an increase in
convective efficiency, or some other effect that increases the
radiation losses at the surface, then the increased cooling would cause the
intergranular downdrafts to be cooler than average (since by this
assumption the granules would have lost more heat). These lower
temperatures would be carried down with the downdrafts, and cause
horizontal average temperatures to be reduced below active regions.
The depth dependence of the effect would depend on the details of the
rate of spreading of the downflows by entrainment.\\

An effect that would cause just such an increased cooling assciated
with active region fields was proposed by Spruit (1977). There it was
shown that the `dimples' in the photosphere caused by the reduced
opacity in the magnetic elements allow more radiation to escape. The
magnetic elements are effectively small leaks through which more heat
escapes than from the normal photosphere.  This increased radiation at
the same time implies a larger average cooling rate in active
regions. The downflows inferred from time-distance helioseismology
(Duvall et al. 1998) are consistent with this interpretation, but are
hard to understand in models with increased sub-surface temperatures.

\newpage

\end{document}